# Data Augmentation and Squeeze-and-Excitation Network on Multiple Dimension for Sound Event Localization and Detection in Real Scenes
## Technical Report

*Byeong-Yun Ko, Hyeonuk Nam, Seong-Hu Kim, Deokki Min, Seung-Deok Choi, Yong-Hwa Park*

Korea Advanced Institute of Science and Technology
Department of Mechanical Engineering, 291 Daehak-ro,
Yuseong-gu, Daejeon 34141, South Korea
{b.y.ko, frednam, seonghu.kim, minducky, haroldchoi6, yhpark}@kaist.ac.kr

**ABSTRACT**

Performance of sound event localization and detection (SELD) in real scenes is limited by small size of SELD dataset, due to difficulty in obtaining sufficient amount of realistic multi-channel audio data recordings with accurate label. We used two main strategies to solve problems arising from the small real SELD dataset. First, we applied various data augmentation methods on all data dimensions: channel, frequency and time. We also propose original data augmentation method named *Moderate Mixup* in order to simulate situations where noise floor or interfering events exist. Second, we applied Squeeze-and-Excitation block on channel and frequency dimensions to efficiently extract feature characteristics. Result of our trained models on the STARSS22 test dataset achieved the best ER, F1, LE, and LR of 0.53, 49.8%, 16.0°, and 56.2% respectively.

***Index Terms***— Sound Event Localization and Detection, Data augmentation, Squeeze-and-Excitation network

## 1. INTRODUCTION

Sound event localization and detection (SELD) is a task detecting desired sound events with their class, timestamps (onset and offset of the sound event), and direction-of-arrival (DoA) [1, 2, 3]. Its application includes surveillance, robotics, monitoring system, etc. Before deep learning based SELD methods, spatial information of sound source was usually predicted by rule-based methods using multi-channel audio signals obtained from microphone array formats such as conventional microphone beamformer, differential and super-directive array, circular/spherical microphone array, binaural microphone array, etc. [4, 5]. However, sound signals obtained in real scenes are distorted by bias error by array platform, reverberation, and background/interference noise. In addition, moving sources degrades the performance of SELD system due to difficulty in recognizing DoA from constantly changing room impulse response (RIR) of sound source. These problems have been limiting performance of rule-based DoA estimation methods [6]. To overcome such problems arising from unpredictable acoustic characteristics in real scenes, data driven methods have been suggested. Especially, deep learning-based methods achieved excellent performance robust to noise and moving source by learning complex pattern of sound source from SELD dataset synthesized by RIRs and target sound event samples [1, 2, 3]. However, to implement SELD on real environments robustly, it should be trained and tested on real SELD dataset, which is recorded on real spatial sound scenes. But obtaining sufficient amount of real SELD data is time consuming and requires extensive works to consider various real environments and precisely label the timestamps and DoA of sound events. Although synthesized dataset has complemented scarce real dataset, dataset characteristics such as sound event occurrences, sound event densities, and amount of polyphony differ from real dataset. In this context, previous deep learning-based methods developed with synthesized dataset tend to be overfitted to synthesized dataset.

We used two main strategies to overcome these problems: 1) applying data augmentation techniques in a way it addresses all dimensions of multi-channel audio data: channel, frequency and time and 2) applying Squeeze-and-Excitation (SE) block on channel and frequency dimension. Data augmentations were applied to Sony-TAu Realistic Spatial Soundscapes 2022 (STARSS22) spatial recording [7] and DCASE2022 synthetic SELD mixtures [8] datasets to prevent overfitting and reduce difference between real and synthesized dataset by transforming acoustic characteristics of sound signal. We used following data augmentation methods: channel swap for channel dimension, pitch shift for frequency dimension, frame shift and time masking for time dimension. In addition, we applied an original data augmentation method proposed in this work, *Moderate Mixup* which further enhance SELD task performance. In addition, we applied SE block on channel and frequency dimensions to apply optimal weighting of channels and frequency components on extracted CNN outputs.

## 2. PROPOSED METHODS

### 2.1. Implementation Details

For input feature, we used spatial cue-augmented log spectrogram (SALSA) consisting of 4 channel log-linear spectrograms and 3 channel eigenvector-based intensity vector [9] on FOA microphone array because it gives additional directional information by approximation of steering vector with eigen principal component analysis. The log-linear spectrogram with 200 frequency bins was extracted using 512 window length and 300 hop length from an audio sample with 24 kHz sampling rate. The sound events and source localizations were recognized using convolutional recurrent neural network (CRNN) architecture with ResNet22 and 2 BiGRU layers [9]. Models presented in this work is trained using both STARSS22 [7] and DCASE2022 synthetic SELD mixtures [8]. For evaluation, we used DCASE2022 Challenge Task3 evaluation metrics [1] consisting of four metrics: location-dependent error rate (ER) and F1 score, class-dependent localization error (LE), and localization recall (LR) in the following analysis. For location-dependent ER and F1 score, a sound event is considered correctly detected only if the class prediction is correct and localization error is less than 20°.

Table 1: A comparison of SELD performance on different output formats. Best score for each metric in each output format is in bold, where the best among all format is in red bold

| Output format | SED threshold | ER | F1(%) | LE(°) | LR(%) |
|---|---|---|---|---|---|
| Decoupled output [9] | 0.3 | 0.75 | **22.3** | **38.8** | **40.9** |
|  | 0.5 | **0.70** | 21.8 | 40.4 | 35.1 |
|  | 0.7 | 0.72 | 21.5 | 42.3 | 32.0 |
| ACCDOA [10] | 0.3 | 0.93 | 25.2 | 37.2 | **56.1** |
|  | 0.5 | 0.64 | **30.4** | **31.4** | 41.8 |
|  | 0.7 | **0.63** | 26.2 | 54.8 | 28.7 |
| Multi-ACCDOA [11] | 0.3 | 0.91 | 22.5 | 41.9 | **58.0** |
|  | 0.5 | 0.68 | **27.8** | **37.5** | 44.4 |
|  | 0.7 | **0.65** | 27.0 | 48.0 | 31.1 |

### 2.2. SELD Output Formats

We compared SELD performance on STARSS22 test dataset depending on output formats: decoupled SELD output (performing event detection and localization separately) [1, 9], ACCDOA [10], and multi-ACCDOA [11] in Table 1. For each format, we applied sound event detection (SED) threshold value of 0.3, 0.5 and 0.7. SED threshold is a criterion that determines if an event is active or not, by applying the threshold value on SED prediction for decoupled SELD output and length of Cartesian DoA vector for ACCDOA and multi-ACCDOA. We observed that optimal SELD performance is achieved on SED threshold of 0.3 for decoupled SELD output and 0.5 for ACCDOA and multi-ACCDOA respectively, considering the number of best performance metrics on each output format. While multi-ACCDOA is supposed to perform better than ACCDOA by increased prediction capability on detection of overlapping events from the same class [11], it performed worse. This seems to be because of larger output dimension of multi-ACCDOA to consider overlapping sound events from same class, which induced underfitting to the small real recording dataset. Based on these results, we applied ACCDOA on the output prediction format with SED threshold of 0.5.

### 2.3. Data Augmentation Methods on Multiple Dimensions

As mentioned in 2.1, we used external data (DCASE2022 synthetic SELD mixtures [8]]), generated by convolving target sound events from FSD50K [12] with measured room impulse responses of Eigenmike in various acoustic scenes [13], to supplement sparse real recording data of STARSS22. Synthetic dataset which consists of 1200 sound samples, is much larger than STARSS22 with 67 sound samples. To prevent overfitting to much larger synthetic SELD dataset, various data augmentation methods were applied to reduce the difference between the synthetic sounds and the real recording dataset by transforming acoustic characteristics of sound signal. We applied data augmentation methods to SALSA on FOA microphone array verified in [9] which are channel swap (CS) [14] and pitch shift (PS, which is the same as frequency shift in [9]). These two data augmentation methods make transformations on channel and frequency dimension of input feature respectively. Dimension of microphone channel is important as it gives physical information regarding the direction of sound events [4]. Frequency dimension affects performance of SED when appropriately handled [15, 16, 17]. In addition, time dimension should also be addressed for SELD because temporal information is crucial in audio domain [18, 19, 20, 21] which handles time-series data. Thus, we applied following data augmentation methods which are shown to enhance SED performance: frame shift (FS) which shifts the input feature and label along time axis [15], and time masking (TM) which masks the random time region of input feature and label [22]. Time masking involves a parameter 'mask ratio' randomly drawn from mask ratio range (set as hyperparameter).

Table 2: A comparison of SELD task performance depending on various data augmentations and combinations. CS, PS, FS, TM, and MM refers to channel swap, pitch shift, frame shift, time masking, and moderate mixup respectively. Best score for each metric in each data augmentation method is in bold, where the best among all data augmentation method is in red bold

| Methods | ER | F1(%) | LE(°) | LR(%) |
|---|---|---|---|---|
| Baseline (CS with probability of 0.5 + PS with frequency shift range of 10) | 0.61 | 39.8 | 20.1 | 53.3 |
| FS with probability of 0.3 | 0.62 | 36.1 | **19.8** | 52.1 |
| FS with probability of 0.5 | **0.61** | **41.0** | 20.0 | 57.8 |
| FS with probability of 0.7 | 0.63 | 38.8 | 23.2 | **58.2** |
| FS with probability of 0.9 | 0.59 | 39.4 | 25.5 | 51.2 |
| TM with mask ratio from 1/6 to 1/10 | 0.61 | 39.3 | 19.7 | 52.5 |
| TM with mask ratio from 1/8 to 1/15 | **0.60** | 37.1 | 19.8 | **55.2** |
| TM with mask ratio from 1/10 to 1/20 | 0.62 | **40.1** | **19.0** | 55.0 |
| TM with mask ratio from 1/12 to 1/25 | 0.61 | 38.1 | 20.6 | 52.1 |
| MM with probability of 0.3 | 0.60 | 38.2 | 22.0 | 55.8 |
| MM with probability of 0.5 | **0.58** | 42.6 | **18.9** | **57.4** |
| MM with probability of 0.7 | 0.61 | **43.2** | 22.8 | 52.5 |
| MM with probability of 0.9 | 0.59 | 40.2 | 20.9 | 50.6 |
| best FS + best TM | 0.60 | 42.2 | 22.0 | 55.3 |
| best FS + best MM | **0.57** | 43.6 | 19.2 | **58.8** |
| best TM + best MM | 0.58 | **43.8** | **18.4** | 55.6 |
| bset FS + best TM + best MM | 0.61 | 42.8 | 20.9 | 54.1 |

In addition, we applied originally proposed data augmentation method named moderate mixup. Moderate mixup is ACCDOA format-based data augmentation method designed to simulate environments where interferences and background noises contaminate target sound events. In ACCDOA format, an active sound event is expressed by a unit vector pointing towards the location of the sound source. Applying mixup [23] directly on ACCDOA unit vectors for two separate sound events with same class would result in label pointing toward direction between the locations of sound sources corresponding to the sound event, where no sound source actually would present. To avoid such distortion of DoA, moderate mixup mixes input features with a mixing ratio selected from beta distribution so that mixing ratio is close to 0 and 1 mostly and the direction information will be represented by the dominant sound source with mixing ratio closer to 1 by taking ACCDOA unit vector of the dominant sound source. As a result, by only mixing input feature and not mixing label. Mixed data simulates interferences and background noises.

Table 3: A comparison of SELD task performance depending on network models. SE, CS, PS, FS, TM, and MM stands for SE block, pitch shift, frame shift, time masking, and moderate mixup respectively.

| Methods | ER | F1(%) | LE(°) | LR(%) |
|---|---|---|---|---|
| no SE with CS + PS | 0.61 | 39.8 | 20.1 | 53.3 |
| Original SE with CS + PS | 0.58 | 40.5 | 19.6 | 52.1 |
| Original SE with CS + PS + FS + MM | **0.56** | 43.6 | **17.1** | **59.1** |
| Original SE with CS + PS + TM + MM | 0.58 | 40.6 | 17.5 | 51.7 |
| Multi-dimensional SE with CS + PS | 0.59 | 41.1 | 19.0 | 55.8 |
| Multi-dimensional SE with CS + PS + FS + MM | 0.59 | 42.3 | 18.5 | 58.3 |
| Multi-dimensional SE with CS + PS + TM + MM | 0.58 | **44.5** | 17.9 | 58.4 |

We experimented on the data augmentation methods at time dimension and moderate mixup to validate each method for SELD and figure out optimal combinations of data augmentation methods. The results are shown in Table 2. Baseline refers to CRNN baseline model with ACCDOA as illustrated in 2.1. The baseline includes channel swap (probability of 0.5) and pitch shift (frequency shift range of 10) as default, as they are already proven in [9]. It is shown that frame shift and time masking, data augmentation methods on time dimension, enhance F1, LE, and LR. In addition, moderate mixup largely enhances F1, and LE from baseline. For the experiment on combinations of data augmentation methods, we observed that applying one of time dimension data augmentation methods with moderate mixup results in best performances, while applying both results in performance degradation from excessive data transformation.

### 2.4. Squeeze-and-Excitation Blocks on Multiple Dimensions

There are 13 sound event classes to detect on DCASE 2022 Task3, including various sound events from harmonic sounds (e.g. human speech and music) to impulse-like sounds (e.g. clapping and door open/close). In addition, the effect of moving source changes acoustic characteristics of sound source constantly. Such harsh conditions make it hard to conduct event detection and predict localization information. Therefore, SELD system should accurately extract features of each sound events from varying complex feature maps. Therefore, we applied Squeeze-and-Excitation (SE) block [24] on ResNet22 structure to adaptively adjust channel weights depending on convolution input with minimal increase of model size. In addition, we applied SE block on frequency dimension as well on each time frame of input to adaptively adjust frequency weights as well.

We compared performances of trained models, and the results are described in Table 3. From 2nd to 4th rows, original SE block which applies on channel dimension is applied. From 5th to 7th rows, multi-dimensional SE block which applied SE on frequency dimension and then on channel dimension is applied. It is shown that applying original SE block on network enhances ER, F1, and LE. Moreover, when utilizing additional data augmentation methods, frame shift and moderate mixup, to the model, LR increases to 59.1. We also observed that applying time masking and moderate mixup with multi-dimension SE block, F1 score largely increases compared to applying frame shift and moderate mixup on the model.

Table 4: Final results of the models submitted.

| Models | ER | F1(%) | LE(°) | LR(%) |
|---|---|---|---|---|
| DCASE2022 Task3 baseline [25] | 0.71 | 21% | 29.3 | 46 |
| (no SE with CS + PS + FS + MM)×2 (Original SE with CS + PS + FS + MM)×3 (Multi-dimensional SE with CS + PS + FS + MM)×3 (Multi-dimensional SE with CS + PS + FS + MM)×3 | **0.53** | **49.8** | **16.0** | 55.9 |
| (Original SE with CS + PS + FS + MM)×3 | 0.55 | 46.2 | 16.4 | 54.6 |
| (Multi-dimensional SE with CS + PS + FS + MM)×3 | 0.57 | 46.4 | 17.2 | 54.4 |
| (Multi-dimensional SE with CS + PS + TM + MM)×3 | 0.55 | 46.4 | 17.0 | **56.2** |

### 3. RESULTS

The results of submitted models for DCASE 2022 Task3 on the STARS22 test dataset are shown in Table 4. Baseline in Table 4 refers to DCASE 2022 Challenge Task3 baseline, different from baseline in Table 2. Model 1 is ensemble average of the 11 models trained with all technics proposed in this paper. Model 2 is an ensemble average on the 3 models trained with original SE block and all data augmentation except time masking. Model 3 is ensemble average on the 3 models trained with multi-dimensional SE block and all data augmentation except time masking. Model 4 is an ensemble average on 3 models trained with multi-dimensional SE block and all data augmentation except frame shift.

### 4. ACKNOWLEDGMENT

This work was supported by "Human Resources Program in Energy Technology" of the Korea Institute of Energy Technology Evaluation and Planning (KETEP), granted financial resource from the Ministry of Trade, Industry & Energy, Republic of Korea. (No. 20204030200050)